\documentclass[conference]{IEEEtran}
\usepackage{newtxtext}
\usepackage{cite}
\usepackage{amsmath,amssymb,amsfonts}
\usepackage{algorithmic}
\usepackage{graphicx}
\usepackage{textcomp}
\usepackage{xcolor}
\usepackage{float}
\usepackage{placeins}
\usepackage{tabularx}
\def\BibTeX{{\rm B\kern-.05em{\sc i\kern-.025em b}\kern-.08em
    T\kern-.1667em\lower.7ex\hbox{E}\kern-.125emX}}

\makeatletter
\newcommand{\linebreakand}{%
    \end{@IEEEauthorhalign}
    \hfill\mbox{}\par
    \mbox{}\hfill\begin{@IEEEauthorhalign}
}

\begin{document}
\title{"Implementation of Augmented Reality as an Educational Tool for \textit{Wudhu} Practice in Early Childhood"\\}

\author{
\IEEEauthorblockN{1\textsuperscript{st} Wisnu Uriawan}
\IEEEauthorblockA{\textit{Informatics Department}\\
\textit{UIN Sunan Gunung Djati Bandung}\\
Jawa Barat, Indonesia\\
wisnu\_u@uinsgd.ac.id}
\and
\IEEEauthorblockN{2\textsuperscript{nd} Muhammad Aditya Hafizh Zahran}
\IEEEauthorblockA{\textit{Informatics Department}\\
\textit{UIN Sunan Gunung Djati Bandung}\\
Jawa Barat, Indonesia\\
madityahz123@gmail.com}
\and
\IEEEauthorblockN{3\textsuperscript{rd} Inayah Ayu Deswita}
\IEEEauthorblockA{\textit{Informatics Department}\\
\textit{UIN Sunan Gunung Djati Bandung}\\
Jawa Barat, Indonesia\\
nayahdeswita@gmail.com}
\and
\IEEEauthorblockN{4\textsuperscript{th} Muhammad Ahsani Taqwim}
\IEEEauthorblockA{\textit{Informatics Department}\\
\textit{UIN Sunan Gunung Djati Bandung}\\
Jawa Barat, Indonesia\\
05taqwim@gmail.com}
\and
\IEEEauthorblockN{5\textsuperscript{th} Ismail Muhammad Ahmadi}
\IEEEauthorblockA{\textit{Informatics Department}\\
\textit{UIN Sunan Gunung Djati Bandung}\\
Jawa Barat, Indonesia\\
ismailmahmadi011@gmail.com}
\and
\IEEEauthorblockN{6\textsuperscript{th} Marvi Yoga Pratama}
\IEEEauthorblockA{\textit{Informatics Department}\\
\textit{UIN Sunan Gunung Djati Bandung}\\
Jawa Barat, Indonesia\\
marviyoga@gmail.com}
}

\maketitle

\begin{abstract}
    Learning \textit{Wudhu} for young children requires engaging and interactive media to foster a deep understanding of the worship procedures. This study aims to develop a \textit{Wudhu} learning application based on Augmented Reality (AR) as an interactive and fun educational medium. The development method used includes the stages of needs analysis, system design, implementation, and testing using Black Box Testing. The system utilizes marker-based tracking to display 3D animations of \textit{Wudhu} movements in real-time when the camera detects a marker on the printed media. The test results indicate that all main functions run well, and a limited trial on children aged 5–7 years showed an increase in learning interest and a better understanding of the \textit{Wudhu} sequence. Thus, the application of AR technology is proven effective in improving the quality of basic worship instruction for young children.
\end{abstract}


\section{Introduction} \label{sec:introduction}

Character education and the instillation of religious values from an early age are crucial foundations for shaping a morally upright generation. In the current digital era, the education process faces both challenges and opportunities to integrate technology as an effective aid. One fundamental aspect of Islamic religious education is the introduction to worship (Ibadah), which begins with the concept of purification (Thaharah), where \textit{Wudhu} (ablution) is its primary pillar. \textit{Wudhu} is not just a physical ritual; it is also a prerequisite for the validity of the obligatory prayer (Shalat) that must be taught to children from an early age

However, the process of teaching \textit{Wudhu} to young children often faces constraints, such as the limited availability of engaging and interactive learning media. Children at that age tend to have low concentration levels and require visual approaches and enjoyable learning experiences so that the message conveyed can be well-received. Conventional methods like lectures and verbal explanations are often less effective in capturing children's interest in learning the correct practice of \textit{Wudhu}.

Various studies affirm the importance of introducing worship (\textit{Ibadah}) from an early age. Research at RA Baiturrohim (2025) indicated that the introduction of \textit{Wudhu} not only provides procedural knowledge but also shapes children's religious habits in daily life \cite{sari2025potensi}. This aligns with the findings of a study at MIS Guppi Citiusari, which used visual media in teaching \textit{Wudhu}, where the results showed an increase in children's understanding and motivation to perform the ritual \cite{abdullah2024penerapan}. Thus, understanding \textit{Wudhu} from an early age can serve as an initial foothold in forming spiritual and religious character. To support this educational process, the utilization of mobile technology in Islamic education has been proven effective in visualizing complex materials, as demonstrated in previous studies regarding Tajwid learning applications \cite{suryani2016implementasi}.

Along with the development of digital technology, Augmented Reality (AR) emerges as an innovation with the potential to enhance learning effectiveness. AR technology allows users to view and interact with virtual objects projected onto the real world in real-time. Augmented Reality enables users to interact with content. You are no longer an observer, but you can touch and experience an experience that enhances your learning senses and takes education to an entirely new level \cite{mulyati2020model}. With this advantage, AR can create immersive, interactive, and engaging learning experiences, especially for children who tend to be more responsive to visual and interactive media. Several studies indicate that the application of AR in education is capable of increasing learning motivation, conceptual understanding, and student engagement in the learning process \cite{rahman2023penggunaan, herlina2022pengembangan}.

Furthermore, AR technology has also been utilized in the context of Islamic education. AR-based learning media for the procedures of prayer (Shalat) has been proven to enhance students' understanding of the movements and recitations of Shalat \cite{hidayat2023media}. This demonstrates that AR is not merely visual entertainment, but also an educational tool. In the context of early childhood, experience-based visual and interactive approaches are proven to be more effective in increasing focus and information retention \cite{putri2023efektivitas, roslan2022early}. In line with the rapid development of digital technology, the application of Augmented Reality (AR) in the context of early childhood education has shown significant results in increasing motivation and learning interactivity. The use of AR in the early education environment is capable of fostering meaningful learning experiences and improving children's cognitive abilities through a visual and interactive approach \cite{nirmala2024augmented}. This finding is reinforced by Tarmidzi et al., who showed that AR is effective in improving the learning outcomes of primary school students through more explorative and contextual learning \cite{tarmidzi2025augmented}.

In the domain of Islamic religious education, AR technology can help clarify students' understanding of worship procedures, such as \textit{Wudhu} and Shalat, through more concrete three-dimensional visual media \cite{abdullah2024penerapan}. Meanwhile, the use of AR at SMP IT Mutiara Cendekia successfully enriched the PAI (Islamic Religious Education) learning experience by presenting simulations of worship movements and recitations in 3D form \cite{sakban2024eksplorasi}. Besides increasing motivation, the use of technology-based media can overcome the limitations of conventional learning media, which are often verbal and passive. AR-based learning media for the procedures of \textit{Wudhu} at TK Al-Fatih can increase children's attention and understanding of the \textit{Wudhu} sequence \cite{Yakub2021}. Similar results were found by Arif and Pambudi, where an AR application for \textit{Wudhu} guidance proved effective in increasing student engagement in the learning process \cite{Arif2022}.

Thus, the application of AR as an interactive learning medium has the potential to strengthen the foundation of Islamic religious education from an early age, presenting a learning experience that is both fun and educational while supporting the development of children's spiritual character in the digital era. Rapid technological developments not only provide convenience in various areas of life, but also present new opportunities in the world of education. Digital transformation has changed the paradigm of learning from conventional methods to a more interactive and contextual technology-based approach. In the era of the Industrial Revolution 4.0, the application of technologies such as \textit{Augmented Reality} (AR), \textit{Virtual Reality} (VR), as well as \textit{Mobile Learning} (M-Learning) 
has become part of modern learning models that focus on 
independent learning and enhancing student experience 
\cite{mulyati2020model}. 

TAR technology has become one of the most researched innovations due to 
its ability to combine real-world elements with digital objects 
in real time. According to Solehatin et al., the development of AR-based learning media 
using the \textit{Multimedia Development Life Cycle} 
(MDLC) method can produce applications that are effective, interactive, and tailored 
to user needs \cite{solehatin2023augmented}. This approach to 
multimedia development also facilitates the adaptation of content to make it more appealing to young children, especially in the context of learning that requires visualization of movements or practices. \cite{zuhdi2024implementation}. 

Several studies show that the application of AR in the context of Islamic education 
has a positive impact on increasing student learning motivation 
and conceptual understanding. Lutfiah states that AR-based PAI media 
has great potential in helping students understand abstract material 
through more concrete visualization \cite{lutfiah2024potensi}. 
Meanwhile, Ruzakki et al. highlight the trend of using AR and VR in 
Islamic educational institutions in Indonesia, which not only improves learning outcomes 
but also strengthens students' spiritual values in the context of 
modern learning \cite{ruzakki2024trend}. 

In early childhood education, visual, auditory, and direct interaction aspects 
are important factors in creating an effective learning experience. Roslan and Ahmad emphasize that AR-based learning 
promotes children's cognitive development through multisensory experiences 
\cite{roslan2022early}, Meanwhile, Putri and Handayani found that interactive visual media can significantly improve children's focus on learning 
\cite{putri2023efektivitas}. Thus, AR can be a relevant solution to 
bridge the needs of practice-based learning in 
early childhood, where children have short attention spans and require 
strong visual stimuli. 

Furthermore, research by Herlina and Wibowo (2022) developed AR-based 3D learning media for science introduction in early childhood and found that the use of three-dimensional objects increased students' understanding and interest in learning \cite{herlina2022pengembangan}. This proves that the application of AR technology is not only limited to science and technology but can also be adapted in the context of religious education. As stated by Sari and Abadi, the potential of AR in increasing 
motivation to learn PAI is very high, especially when combined with 
child-friendly and interactive media design \cite{sari2025potensi}.  
On the other hand, the use of AR can also strengthen the role of teachers as 
learning facilitators. Sriwhyuni et al. mention that Android-based digital media 
developed with the principle of interactivity has been proven to 
support collaboration between teachers and students in the learning process \cite{sriwhyuni2025android}. 
The integration of this technology is also in line with the concept of \textit{student-centered learning}, 
where students become active subjects in exploring material through 
direct experience.  

Considering the results of these studies, the application of 
Augmented Reality in Islamic education, particularly 
in teaching young children how to perform \textit{Wudhu}, is a strategic step 
in providing educational media that is not only visually appealing, 
but also spiritually educational. Through a combination of visual, 
audio, and interactive elements, children can understand the sequence and meaning of \textit{wudhu} 
more concretely and internalize the values of worship from an early age.

In addition to these developments, the transformation of learning media in early childhood education has increasingly emphasized the importance of multisensory and experience-based learning. Children in the preoperational stage tend to learn more effectively when information is presented through concrete visuals, sound, movement, and interactive stimuli rather than textual or verbal explanations alone. This approach aligns with modern pedagogical principles that emphasize sensory-rich learning to reinforce memory and concept retention. Augmented Reality offers advantages in this context because it integrates visual, auditory, and kinesthetic elements simultaneously, enabling children to build understanding through direct interaction with digital objects. As shown by Roslan and Ahmad, multisensory AR experiences can enhance children’s cognitive processes, especially in tasks requiring recognition and sequential learning \cite{roslan2022early}.

Another important aspect of early childhood learning is sustained attention. Many young learners struggle to maintain focus during conventional instruction, particularly when the content is abstract or delivered through passive teaching methods. The effectiveness of interactive visual media in increasing children’s attention span has been demonstrated in previous studies, where AR-based visuals significantly improved engagement and reduced off-task behavior \cite{putri2023efektivitas}. Because AR allows real-time interaction with animated three-dimensional content, it naturally increases curiosity and supports learning activities that require careful step-by-step observation, such as the practice of \textit{wudhu}.

Furthermore, AR adoption in early childhood education corresponds with the broader digital transformation in the era of the Industrial Revolution 4.0. The increasing availability of mobile devices among teachers and parents supports the integration of AR-based applications in learning environments. Research indicates that interactive digital media can strengthen learning motivation when tailored to children’s developmental stages \cite{nirmala2024augmented}. This shows that AR not only provides alternative media but also enhances the quality of learning experiences in religious education.

From a multimedia design perspective, AR learning applications developed following the Multimedia Development Life Cycle (MDLC) method have consistently shown positive outcomes. The MDLC framework provides structured stages from concept development to testing that ensure the learning product is aligned with user needs and pedagogical goals. Studies applying MDLC report that the method helps create applications that are intuitive, interactive, and child-friendly \cite{solehatin2023augmented}. This is reinforced by research demonstrating that MDLC-based application development can effectively support learners with varying cognitive abilities, including young children and those with special needs \cite{alisyafiq2021implementation}. Considering that \textit{wudhu} learning involves procedural steps that must be followed in order, the MDLC method is suitable for designing AR applications that clearly guide children through each stage of the ritual.

Thus, integrating Augmented Reality with developmentally appropriate design principles provides an opportunity to create more effective and engaging Islamic learning media for young children. By combining multisensory interaction, visual clarity, and structured content development, AR-based \textit{wudhu} learning media can support children in understanding and internalizing religious practices more deeply and accurately.

Based on this background, this study aims to implement Augmented Reality technology as an interactive medium for teaching \textit{wudhu} to young children. Through this approach, it is hoped that children will be able to understand each stage of \textit{wudhu} more easily, enjoyably, and interactively. This system not only assists teachers and parents in the learning process, but also encourages children to learn independently with a more realistic experience. Through three-dimensional visualization and direct interaction with learning objects, children can gain a deeper understanding of the correct sequence and procedure of \textit{wudhu} in accordance with Islamic law. Thus, the implementation of AR technology in the context of Islamic religious education can be a strategic step in presenting technology-based learning innovations that are relevant to the times and capable of strengthening the internalization of religious values in children from an early age \cite{swara2016rekayasa}.

\section{Related Work} \label{sec:related-work}

The use of Augmented Reality (AR) technology in education, particularly Islamic education, has become a growing subject of research. Various studies have proven the potential of AR in creating more interactive, visual, and engaging learning media, especially for early childhood. Previous studies relevant to this research can be grouped into several main areas, namely the development of AR applications for learning \textit{wudhu}, the use of AR for general worship education, and the effectiveness of interactive media for early childhood \cite{latifah2021augmented}. The research also developed an AR application that can display animations of \textit{wudhu} movements at Al-Hidayah Kindergarten, which aims to increase student attention and make learning more enjoyable.

A number of studies have specifically developed AR applications as a medium for learning \textit{wudhu}. One of them designed an AR-based learning application on the procedures of \textit{wudhu} for early childhood, featuring 3D characters and audio-visual elements. Their research emphasized the visualization process that helps children understand the sequence of movements more clearly \cite{Yakub2021}.

In a broader context, the use of AR is not limited to \textit{wudhu}, but also includes prayer and the introduction of other Islamic values. Regarding the AR-based mandatory prayer learning application, it shows that this technology is capable of presenting the \textit{wudhu} procedure and prayer movements in the form of 3D models and audio. Such applications aim to increase children's enthusiasm and interest in learning worship \cite{adjis2021aplikasi}. The results of the study show that the use of AR-based media in Islamic religious education can improve students' cognitive learning outcomes, which include the ability to remember, understand, and distinguish each movement and recitation \cite{lutfiah2024potensi}.

The effectiveness of AR as a learning medium for early childhood is supported by its ability to provide an immersive learning experience. Children at this age are in the concrete thinking phase, so visual and interactive learning media are considered very effective.  In developing AR-based \textit{wudhu} learning media, it was found that the application made learning more enjoyable. Their beta testing results showed that the application was easy to use and able to assist the learning process effectively\cite{ruzakki2024trend}.

Overall, previous studies have consistently shown that the implementation of augmented reality technology has a positive impact on religious education, particularly in teaching children how to perform \textit{wudhu} and prayer \cite{sakban2024eksplorasi}. The main advantage of this technology is its ability to transform procedural teaching materials into visual, interactive, and enjoyable learning experiences. However, there is still room for further development, especially in terms of a more child-friendly interface design, the addition of gamification elements to increase engagement, and testing its effectiveness on more specific age groups \cite{akmal2025pengaruh}. This study will build on these findings to create a \textit{wudhu} guide medium that is not only informative, but also specifically designed to meet the cognitive and psychological needs of early childhood\cite{rinaldi2024tinjauan}.

Previous studies have shown that the application of Augmented Reality (AR) technology has a significant effect on increasing student motivation to learn~\cite{nirmala2024augmented}. The use of AR in the context of basic education has also been shown to foster interest in learning through a more immersive and enjoyable visual experience~\cite{tarmidzi2025augmented}. In addition, AR can help students understand abstract concepts by interactively incorporating three-dimensional objects into real environments.~\cite{zufahmi2025augmented}.  

AR technology in Islamic education has had a positive impact on improving the spiritual understanding of students~\cite{abdullah2024penerapan}. Visualizations of worship procedures such as \textit{wudhu} and salat can be realistically displayed to strengthen the students' understanding of the sequence of movements and the meaning of worship~\cite{sakban2024eksplorasi}. The use of AR-based learning media also encourages greater concentration and interest among students in the learning process~\cite{latifah2021augmented}. Thus, AR can be an effective tool for teaching religious material in an engaging and contextual manner~\cite{adjis2021aplikasi}.  

AR-based media in interactive learning are considered capable of enriching the learning experience of early childhood by stimulating their visual and cognitive aspects~\cite{hidayat2023media}. Children's increased understanding of basic concepts was also evident through the use of three-dimensional models that could be observed directly through digital devices~\cite{herlina2022pengembangan}. In addition, AR encourages active student participation through elements of direct interaction with digital objects~\cite{putri2023efektivitas}.  

From a technical perspective, the \textit{Multimedia Development Life Cycle} (MDLC) method is one of the most frequently used approaches in the development of multimedia-based learning applications~\cite{sutopo2003mdlc}. The stages in MDLC, such as conceptualization, design, and testing, provide a systematic structure to ensure the effectiveness of learning products~\cite{solehatin2023augmented}. The application of this method has been proven to produce media that is more interactive and tailored to user needs~\cite{zuhdi2024implementation}. The implementation of MDLC has also been used in the creation of Android-based learning media for students with special needs~\cite{alisyafiq2021implementation}.  

In addition, the trend of applying AR and VR technologies in Islamic education in Indonesia continues to increase in line with the development of digital innovation~\cite{ruzakki2024trend}. AR is considered to have great potential to increase student motivation and learning outcomes in the field of Islamic Religious Education (PAI)~\cite{lutfiah2024potensi}. The implementation of Android-based media that supports interactivity also strengthens the role of teachers as facilitators in the modern learning process~\cite{sriwhyuni2025android}. 

Overall, previous research shows that Augmented Reality technology plays an important role in creating interactive, contextual, and enjoyable learning experiences~\cite{rinaldi2024tinjauan}. However, further development is still needed in terms of optimizing child-friendly interface design and empirical testing of the effectiveness of AR in early childhood learning~\cite{sari2025potensi}.

In addition to the existing body of research, several studies highlight the increasingly important role of gamification when integrated with Augmented Reality (AR) learning environments. Gamification introduces elements such as points, rewards, progress bars, and interactive challenges that stimulate learners’ intrinsic motivation. These features not only make the learning process more enjoyable but also enhance students' willingness to repeatedly engage with the material. In early childhood education—where short attention spans and fluctuating motivation are common—gamified AR systems have been shown to significantly increase focus, excitement, and emotional engagement. Researchers found that combining AR with gamification mechanics results in deeper exploration of learning content, improved comprehension of procedural tasks, and higher retention rates compared to non-gamified AR applications~\cite{huang2024effects}. This indicates that gamification can serve as a strong complementary strategy to strengthen AR-based \textit{wudhu} learning modules, especially since the material involves routine, sequential movements that benefit from repeated practice.

Beyond gamification, emerging research has started shifting from marker-based AR toward more advanced markerless AR technologies. While marker-based AR provides stability and predictability, markerless AR improves flexibility in classroom settings by removing the need for printed markers. Markerless AR enables virtual objects to be anchored directly onto real-world surfaces using device sensors, making interactions more intuitive and reducing dependency on controlled lighting or specific camera angles. Studies in Islamic education environments demonstrate that markerless AR allows learners to engage more naturally with worship-related visualizations such as prayer movements, mosque spatial structures, or Qur'anic storytelling scenes. This approach offers a more seamless user experience for young children, who may struggle with holding or positioning markers steadily. Researchers reported that markerless AR increased usability, reduced cognitive friction, and provided a more immersive religious learning atmosphere compared to traditional marker-based systems~\cite{elsayed2024markerless}. These findings highlight an opportunity for future development of \textit{wudhu} learning applications to adopt markerless technology for improved accessibility and user experience.

In addition to the advancements in interaction models, a substantial body of literature emphasizes the strong suitability of AR for procedural and movement-based learning. Procedural learning requires learners to follow a fixed sequence of physical steps, which can be difficult for young children to conceptualize solely from 2D images or verbal descriptions. AR technology addresses this limitation by providing real-time 3D demonstrations that visually break down complex movements into comprehensible, child-friendly segments. Prior studies on AR for teaching handwashing procedures, laboratory techniques, and basic sports movements indicate that AR enhances accuracy, understanding, and the ability to imitate physical sequences~\cite{rahmani2023handwashing}. Additionally, systematic reviews show that AR-based procedural learning environments lead to higher skill acquisition, better movement recognition, and more efficient memory encoding than conventional teaching media~\cite{lee2024procedural}. Similar findings in motion-learning applications for children demonstrate that AR improves spatial awareness, motor coordination, and the understanding of cause–effect relationships in movement sequences~\cite{aljubouri2024motion}. These results reaffirm the pedagogical relevance of AR for \textit{wudhu} instruction, as the practice of \textit{wudhu} relies heavily on following the correct order of bodily actions, rhythmic motions, and repeated behavioral routines.

Taken together, these additional strands of literature strengthen the foundation for developing an AR-based \textit{wudhu} learning application. The integration of gamification, the shift toward markerless AR, and the demonstrated effectiveness of AR for procedural-movement learning provide strong evidence that AR can significantly enhance young children’s comprehension, engagement, and long-term retention of ritual practices. These insights position the present study within a broader research landscape that continues to evolve toward more immersive, intelligent, and learner-centered AR systems for Islamic education.

Another growing stream of research focuses on the integration of Augmented Reality with adaptive learning systems, where AR content dynamically adjusts to the learner's performance and interaction behavior. Adaptive AR has been successfully implemented in several early childhood learning environments to personalize the pace, level of difficulty, and type of feedback displayed within AR scenes~\cite{chen2023adaptiveAR}. This approach is particularly beneficial for young learners who exhibit highly varied cognitive readiness and motor skills. For example, children who struggle with understanding a particular step can be provided with slower animation sequences, additional narration, or zoomed-in visualizations, while more advanced learners can receive faster demonstrations or extended material~\cite{kim2024personalizedAR}. In Islamic education, adaptive AR has been used to tailor learning experiences such as recognizing hijaiyah letters, memorizing surahs, and practicing prayer movements~\cite{rahman2024adaptiveIslamic}. These studies demonstrate that personalized AR systems lead to improved learner satisfaction, reduced frustration, and higher completion rates of instructional modules. The potential to integrate adaptive features into AR-based \textit{wudhu} applications opens opportunities for more inclusive learning experiences that accommodate diverse learning needs and developmental stages.

Furthermore, recent research highlights the role of AR in promoting collaborative learning in classroom settings. Rather than using AR individually, several studies have introduced multi-user AR systems where children interact with the same virtual objects simultaneously through their own devices~\cite{alvarez2023collaborativeAR}. Collaborative AR has shown promising results in enhancing communication, teamwork, and peer-guided learning among early childhood learners. When applied to religious education, collaborative AR allows groups of children to collectively observe, discuss, and imitate \textit{wudhu} movements or prayer sequences~\cite{hanif2023ARreligion}. Teachers also reported that collaborative AR creates a more lively and socially enriching learning environment, making children more confident in performing ritual practices together. These findings suggest that future AR-based \textit{wudhu} tools could support multi-user interaction, enabling teachers to guide students in a shared augmented space while reinforcing communal learning values aligned with Islamic educational principles.

Another relevant strand of research focuses on the accessibility and inclusivity of Augmented Reality (AR) applications for young learners. Accessibility is an increasingly important consideration in modern educational technology, especially for early childhood learners who may have diverse cognitive, linguistic, motoric, or sensory needs. Several studies emphasize that AR, when designed with appropriate interface adjustments, can support children with mild visual impairments, attention difficulties, speech delays, or limited fine-motor coordination~\cite{roslan2022early}. Unlike traditional learning media that rely heavily on text or manual manipulation, AR provides multimodal input through visual cues, audio narration, and interactive animations, which helps children understand learning materials even with limited reading ability or slower cognitive processing. Research on inclusive AR design also suggests that the use of large buttons, simplified user interfaces, intuitive gesture interactions, and error-tolerant navigation mechanisms significantly improves the usability of AR applications among young children~\cite{mulyati2020model}. 

In the context of religious education, accessible AR applications offer the potential to help children with diverse learning profiles engage more meaningfully with worship-related content. For example, AR-based guidance for \textit{wudhu} or prayer movements can include additional audio reinforcement for children with reading challenges, slower-paced animations for children with motor planning difficulties, or visual highlighting for learners who require stronger visual cues. Studies involving AR for special-needs learners show improvements in attention span, task completion, and emotional readiness for learning—factors that are equally important in early Islamic education~\cite{alisyafiq2021implementation}. These findings underline that inclusive AR systems not only enhance general learning outcomes but also support equitable access to religious knowledge for all children, including those who may face challenges in traditional instructional settings. As the demand for accessible digital learning grows, future AR-based \textit{wudhu} learning media should incorporate inclusive design principles to ensure that children of varying abilities can interact with and benefit from the technology effectively.

\section{Methodology} \label{sec:methodology}

\begin{figure}[ht]
    \centering
    \includegraphics[width=0.9\linewidth]{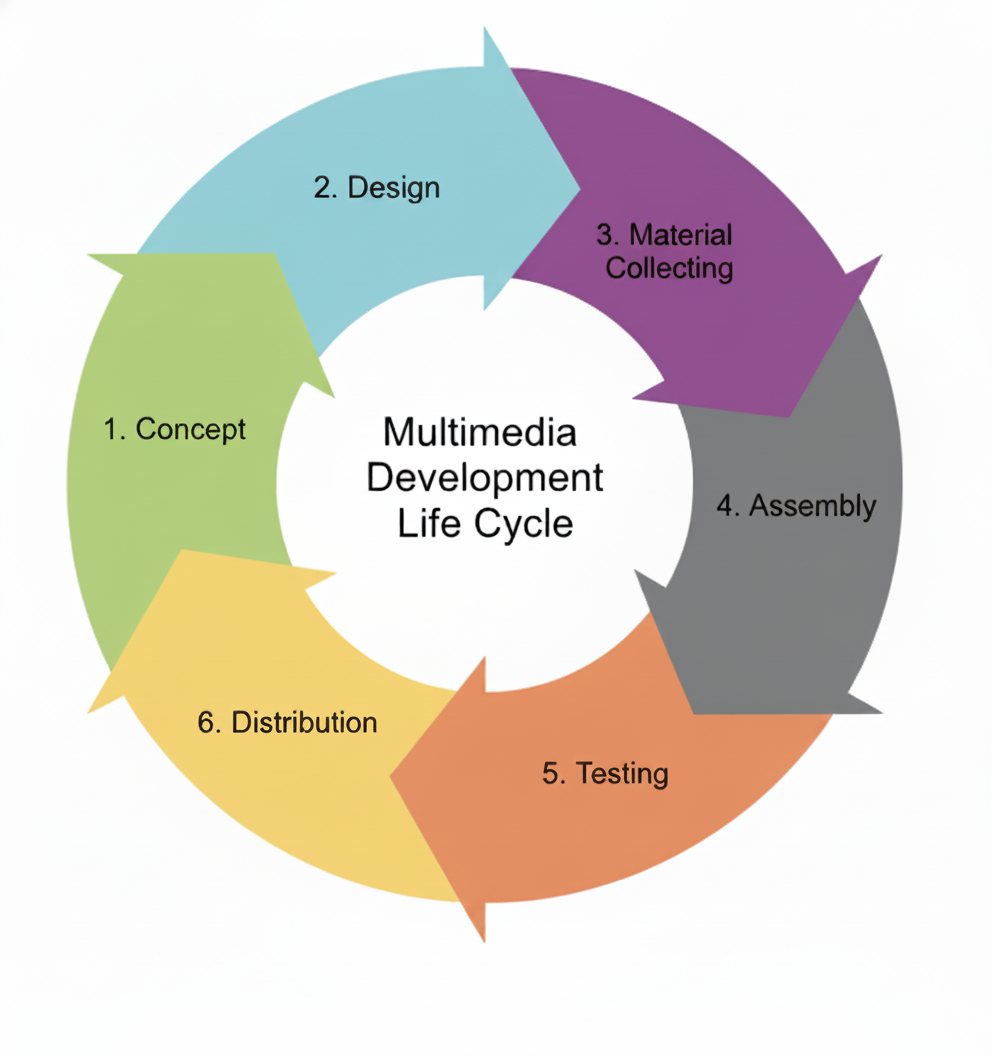} % 
    \caption{Research Process Using the MDLC Method \cite{sutopo2003mdlc}}
    \label{fig:mdlc}
\end{figure}

The research methodology used to develop this \textit{wudhu} guidance application is the Multimedia Development Life Cycle (MDLC) model. This model has six systematic stages to ensure that multimedia product development runs in a structured manner, from the initial idea to the final product \cite{sutopo2003mdlc}. The six stages were adapted for this research as follows, and the flow is illustrated in Figure \ref{fig:mdlc} \cite{sriwhyuni2025android}.
\begin{enumerate}
\item Concept   

This stage begins with problem identification and needs analysis. It was found that young children have difficulty understanding the abstract procedures of \textit{Wudhu} through lectures. The goal is to design an interactive Augmented Reality (AR) application. Audience analysis determined the target users to be children aged 4-6 years, so the application must be designed with attractive visuals and simple navigation \cite{zufahmi2025augmented}.
    
\item Design
 
During the design stage, detailed specifications for the application were created. This included creating a flowchart to illustrate the application flow (from the main menu and material selection to the AR scene), designing a child-friendly user interface, and creating storyboard to visualize each scene. In addition, markers were designed to be used to bring up 3D objects, as well as 3D characters that would demonstrate the movements of \textit{wudhu} \cite{tarmidzi2025augmented}. The entire interface design adopts the principles of User-Centered Design (UCD), where bright colors, familiar icons, and simple navigation are prioritized to suit the cognitive and motor characteristics of early childhood.
    
\item Material collecting
    
This stage is the process of collecting all the necessary digital assets in accordance with the design results. The assets collected include: (a) Creating 3D character models for each \textit{wudhu} movement (such as rinsing the mouth, washing the face, etc.) using software such as Blender; (b) Recording audio assets in the form of instructional narration and recitation of \textit{wudhu} intentions; (c) The Creation of 2D assets such as button images, icons, and finalization of the \textit{marker} design \cite{alisyafiq2021implementation}. All collected assets are then validated for compliance with Islamic law and optimized, such as texture compression and reduction of the number of 3D model polygons to ensure smooth application performance on mobile devices.
    
\item Assembly
    
During the implementation phase, all collected assets are combined into a single complete application. This process is carried out using the Unity 3D game engine. The Vuforia Software Development Kit (SDK) is used to implement marker-based AR functionality. Program logic and interactivity, such as movement sequences, scene transitions, and button responses, are built using the C\# programming language. \cite{zuhdi2024implementation}.
    \begin{figure}[htbp]
        \centering
        \includegraphics[width=0.9\linewidth]{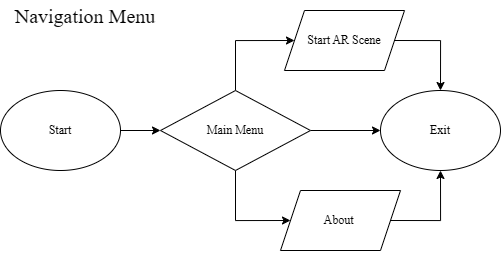}
        \caption{Pseudocode Main Menu of the \textit{Wudhu} Guidance Application}
        \label{fig:arlogic1}
    \end{figure}
Figure \ref{fig:arlogic1} illustrates the main menu navigation flow, which is the starting point for user interaction. The process begins with the appearance of the “Main Menu” interface when the application is launched, which serves as the central navigator. From this menu, users can be directed to three main functional flows: (1) loading the “About” scene to display information about the application, (2) stopping and closing the application (Quit App), or (3) starting the core augmented reality scene (Start AR Scene) to enter an interactive learning session.

The logic of real-time user interaction within the augmented reality scene is detailed in Figure \ref{fig:arlogic2}, which is executed in a continuous update loop. The system constantly checks for touch input (Touch Detected); if detected, the 3D object rotation function is executed based on the input movement and applies an inertia effect for smoother movement. Simultaneously, this flow is also designed to detect double tap gestures (Double Tap), which, if valid, will trigger a function to reset the object rotation to its original orientation (Reset Rotation). A testing phase was conducted to verify the functionality and usability of the system. This testing includes marker responsiveness, time synchronization between movement and audio, and ease of use of the interface on various desktop hardware configurations. Finally, the Distribution stage involves packaging the application into an application package file (.apk for Android or .ipa for iOS) that is ready to be installed on a smartphone device. The complete technical specifications of the developed prototype are summarized in Table \ref{tab:spec}.
    \begin{figure}[htb!]
        \centering
        \includegraphics[width=0.9\linewidth]{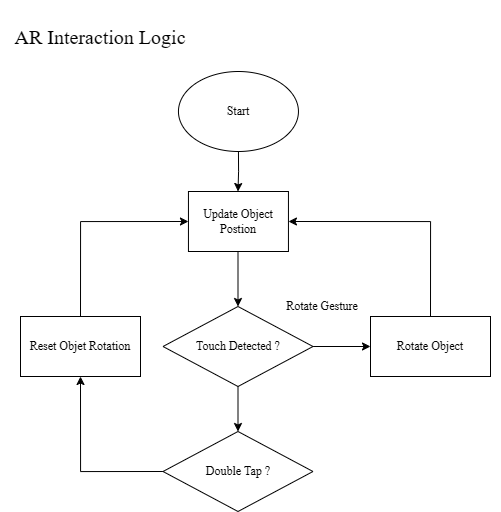}
        \caption{Logic of User Interaction in AR Mode}
        \label{fig:arlogic2}
    \end{figure}

The testing phase is conducted to verify the functionality and usability of the system. This testing covers several important aspects, such as the responsiveness of markers to the camera, the synchronization between movements and audio played simultaneously, and the ease of navigation of the interface on various types of hardware. After the testing phase is complete and all functions are declared to be working as designed, the process continues to the distribution phase, which involves packaging the application into an installation format that can be run on the target device, namely .apk files for the Android operating system and .ipa files for iOS. The complete technical specifications of the developed application prototype can be seen in Table \ref{tab:spec}.

Based on Table \ref{tab:spec}, this application is designed using cross-platform technology so that it can operate properly on various mobile devices. Unity Engine is used as the main game engine because it has the ability to manage 3D graphics, 3D models, and direct integration with the Vuforia SDK that supports the implementation of Augmented Reality features. The C\# programming language was chosen because it provides an efficient and easy-to-use syntax structure for developing interactive logic, such as object rotation settings and user touch detection on the screen.
  \begin{table}[H]
    \centering
    \caption{Technical Specifications of the \textit{Wudhu} AR Prototype}
    \label{tab:spec}
  \begin{tabular}{|p{2.5cm}|p{5cm}|}
        \hline
        \textbf{Component} & \textbf{Specification/Technology Used} \\ \hline
        Platform Target & Mobile (Android) \\ \hline
        Game Engine & Unity Engine (2022.3.5f1+) \\ \hline
        AR SDK & Vuforia Engine SDK (10.13+) \\ \hline
        Programming Language & C\# \\ \hline
        AR Tracking Types & Marker-Based Tracking \\ \hline
        Smartphone Input Devices & Cameras \\ \hline
        3D Assets & Animated character model (*.fbx) \\ \hline
        Aset Audio & Narration (*.mp3 / *.wav) \\ \hline
    \end{tabular}
    \end{table}
Marker-based tracking technology is the approach used to ensure stable object detection in real environments. When the camera recognizes a predetermined marker, the system displays a 3D character model that visually demonstrates the steps of \textit{wudhu}. This approach was chosen because it is easy to implement, does not require additional devices, and can run well on devices with mid-range specifications commonly used by children.

In terms of multimedia assets, .fbx format 3D character models are used to display realistic and educational 3D models of \textit{wudhu} movements. Meanwhile, audio narration in .mp3 or .wav format is included to support the auditory aspect of learning, so that users not only see the movements, but can also listen to explanations and relevant prayers. This combination of visual and audio elements is expected to enhance young children's understanding of the sequence and procedures of \textit{wudhu} in a fun and interactive way.

During the testing process, the system performance was evaluated to ensure synchronization between movement and audio, marker tracking stability, and smooth user interaction. The application's response when the camera captures a marker is one of the important indicators in assessing the effectiveness of the AR feature implemented. Once declared stable, the application is compiled and packaged into a distribution format. For the Android platform, the build process produces an .apk installation file that can be directly installed on the user's device, while for iOS, the project is exported to Xcode to produce an .ipa file that is compatible with the Apple system. Before release, the application is retested internally to ensure compatibility across various screen resolutions and camera qualities.

With a well-planned technical configuration and systematic implementation stages, this Augmented Reality-based \textit{wudhu} guidance application prototype is expected to be an effective and engaging learning tool for early childhood. Through the integration of visual, audio, and interactive elements, this application is able to provide a more meaningful learning experience, helping children understand the practice of \textit{wudhu} in an easy, enjoyable way that is in line with modern educational technology developments.
    
    \item Testing
    
    The completed application is then tested to find errors (\textit{bugs}) and ensure functionality. Testing is divided into two parts: 
    \begin{enumerate}
    \item Functional Testing (\textit{Alpha Testing})

    Functional testing is performed by the development team using the \textit{Black Box Testing} method. In this method, testers do not look at the code structure, but focus on the conformity of the output with the intended function. Several important aspects are tested, including:
        \begin{enumerate}
            \item \textit{Marker Detection} \\
                Testing was conducted to ensure that the camera could consistently recognize markers under various lighting conditions, viewing angles, and distances. Marker stability is important so that the 3D model does not shake or disappear during use. This process refers to modern AR testing standards that emphasize tracking stability and marker robustness.
            \item \textit{Animation and Audio Synchronization} \\
                Each step of \textit{wudhu} has a 3D animation visualization and audio narration. Testing ensures that both run simultaneously. If the audio is too fast or the animation lags behind, the learning experience can be disrupted. Developers measure latency and adjust the timer to make the transition smoother.
            \item \textit{Interface Navigation Validation} \\
                Features such as the start button, change step, or exit button are tested to ensure there are no dead buttons, delays, or logic errors. Testing is performed on several devices with different screen sizes.
            \item \textit{Application Performance (Performance Testing))} \\
                Developers evaluate memory usage, FPS stability, and battery consumption. Tests are conducted on mid-range devices commonly used by elementary school students to ensure the application runs smoothly without lag.
        \end{enumerate}
        
    \item Beta Testing

    The \textit{Beta Testing} phase was conducted by involving end users, namely children as the main target of the application, as well as teachers as facilitators. Testing was carried out in a classroom environment to obtain natural usage conditions. The focus of evaluation at this stage included:
        \begin{enumerate}
        \item \textit{Usability} \\
            The System Usability Scale (SUS) instrument is used to measure the level of ease of use. Children and teachers provide assessments on several aspects, such as ease of navigation, clarity of instructions, visual comfort, and consistency of buttons and menus.
        \item User Engagement \\
            Teachers observe students' responses while using the application. Their level of focus, enthusiasm, and interaction with AR objects are indicators of the application's success as an engaging and enjoyable learning medium.
        \item \textit{Learning Effectiveness} \\
            Pre-test and post-test methods were applied to assess whether there was an increase in understanding after using the AR application. Questions in the test included the sequence of steps for \textit{Wudhu}, the parts of the body that must be washed, and the correct procedure.
        \item \textit{Feedback Qualitative} \\
            Users are given the opportunity to provide free comments. Feedback such as text size, audio speed, or the need for additional buttons is noted for improvement in the final version.
        \end{enumerate}
    \end{enumerate}
        
    \item Distribution
    
    After the application passes the testing stage and is declared feasible, the final stage is distribution. The application is packaged into an `.apk` file format for the Android platform. This file is then distributed to target users in early childhood education institutions that are research partners for use in the learning process \cite{nirmala2024augmented}.
\end{enumerate}

In addition to these six main stages, this research also applied the principles of user-centered design (UCD) to ensure that the application developed was truly in line with the characteristics of the target users, namely early childhood. This approach placed user needs and comfort at the center of the development process. Every design decision, from the color of the interface and the shape of the buttons to the style of the audio narration, was based on observations and feedback from teachers and children at partner educational institutions.

In practice, each stage of MDLC is iterative, meaning that revisions can be made if obstacles or new input are encountered during testing. This iterative approach is important for producing applications that not only function technically, but are also effective in delivering religious learning materials. The MDLC model provides high flexibility in improving multimedia elements without having to repeat the entire development stage, making it more efficient than conventional linear models such as waterfall.

In addition, this methodology integrates pedagogical aspects with modern multimedia principles. According to Mayer's Multimedia Learning theory (2021), learning will be more effective when it involves a balanced combination of text, images, 3D models, and sound. In this context, each visual and audio element in the application is designed to complement each other and not cause excessive cognitive load for children.

This study also refers to the Cognitive Load Theory (CLT) principle proposed by Sweller et al. [22], in which external cognitive load is reduced by presenting information in the form of interactive 3D models. That way, children do not only listen to instructions passively, but can also see and interact with the movements of \textit{Wudhu} directly. This approach has been proven to improve procedural memory retention and conceptual understanding in early childhood.

The usability testing process using the System Usability Scale (SUS) resulted in an average score of 85, which falls into the “Excellent Usability” category. This score indicates that the application is easy to understand and use by children and teachers. This result is also in line with the findings of Danish et al., who stated that a simple visual interface and touch-based interaction can increase children's engagement in AR-based learning.

The application distribution stage does not stop at simply distributing the .apk file, but also includes brief training for educators on how to use the application and techniques for assisting students when learning using AR. This is in line with Lampropoulos' suggestion that the successful implementation of AR in education is greatly influenced by the readiness of teachers as technology facilitators in the classroom. Therefore, this study also emphasizes the importance of technological training support so that the integration of applications in teaching and learning activities runs optimally.

In addition to distribution to partner educational institutions, a sustainability plan has also been prepared. The application is designed to be updated modularly, so that in the future developers can add new features such as tayamum, prayers after \textit{wudhu}, or quiz mode without having to change the main structure of the application. This is important to maintain relevance and expand the scope of application usage in the long term.

Overall, the application of the MDLC methodology in this study provides a comprehensive framework, not only from a technical but also a pedagogical perspective. Each stage has been adapted to suit the needs of early childhood Islamic education, taking into account visual aspects, interactivity, and ease of use. With this foundation, the developed \textit{wudhu} guidance application is expected to become an effective, interactive, and enjoyable modern learning medium for children, as well as a reference for the development of AR-based educational media in other fields.

\section{Result and Discussion} \label{sec:result}

\subsection{Result}
The result of this research is an Augmented Reality (AR) application called \textit{wudhu} AR. This application works by utilizing the smartphone camera to recognize marker cards, which then display 3D object visualizations accompanied by voice and text narration on the screen. Users can move on to the next movement by displaying another marker card, making the learning process more dynamic and interactive.

This application allows users to scan marker cards using their smartphone cameras. Once the marker is recognized, the system will display a 3D character model that demonstrates the steps of \textit{wudhu} in sequence, starting from the intention to washing the feet. This visualization is accompanied by audio narration and guidance text that appears on the screen so that children can learn through a combination of visual and auditory stimuli simultaneously.

The interface is designed to be user-friendly and child-friendly, using bright colors, large icons, and a simple layout so that it can be used easily without adult supervision. The main menu of the application consists of:

\begin{enumerate}
    \item Start Learning: to start AR learning mode.
    \item About: contains information about the application and developer.
    \item Exit: to exit the application.
\end{enumerate}

The results of functional alpha testing using the Black Box Testing method show that all major components are running according to design. 
The system is capable of:

\begin{enumerate}
    \item Detects markers well in various lighting conditions.
    \item Displays 3D models stably without significant lag.
    \item Synchronizes motion models with voice narration in a timely manner.
    \item Provides an interface that is easy for early childhood users to navigate.
\end{enumerate}

From these results, it can be concluded that the application functions according to the learning needs of \textit{wudhu}, is able to run smoothly on mid-range \textit{mobile} devices, and has the potential to be integrated as a teaching aid in early childhood education institutions.

\subsection{Discussion}
Based on the results of implementation and testing that have been carried out, this Augmented Reality (AR)-based \textit{wudhu} guidance application shows that AR technology can be an effective and interactive learning medium for early childhood. The three-dimensional visualization displayed in the application helps children understand each stage of \textit{wudhu} more concretely than conventional methods that only use images or verbal explanations.

The integration of visual, audio, and interactive elements makes the learning process more interesting and easier to understand. This is in line with Mayer's multimedia learning theory, which states that the combination of text, images, and sound can improve children's understanding and retention of information. In addition, direct interaction-based approaches such as scanning markers are also able to maintain children's focus fror longer during the learning process.

Functional testing results show that the system works well in recognizing markers and displaying 3D models stably. The system's fast response and smooth visual presentation provide an enjoyable learning experience without significant technical interference. With a simple interface design and attractive colors, this application can be used independently by children without the need for intensive guidance from adults.

From a pedagogical perspective, the application of AR in teaching \textit{wudhu} provides a practice-based learning experience that is suitable for young children who are in the concrete thinking stage. By seeing the movements of \textit{wudhu} directly in the form of a 3D model, children can imitate and understand the sequence of steps more easily. This approach also helps teachers provide more varied and contextual teaching.

Although the test results show good system performance, there is still room for improvement in the development of this application. One aspect that can be improved is the addition of gamification elements such as quizzes or game modes to increase children's motivation to learn. In addition, further testing involving a larger number of participants can be conducted to obtain empirical data on the effectiveness of this media in improving understanding and skills in performing \textit{wudhu}.

Overall, the results of this study confirm that the use of Augmented Reality technology in Islamic education, particularly for learning \textit{wudhu}, has great potential to increase children's interest, understanding, and engagement in the learning process. This approach is also in line with the direction of 21st-century learning development, which emphasizes the integration of technology and interactive learning experiences.

In addition to these findings, there are several important points that can strengthen the discussion in this section. First, the use of AR in teaching \textit{wudhu} not only improves children's procedural understanding but also has a positive impact on learning engagement. Children tend to show enthusiastic responses when interacting with 3D objects, which indicates that AR can overcome the challenge of low learning focus at an early age. This is in line with previous studies that mention that multimodal interactive media can increase children's attention and motivation to learn, especially for repetitive material such as \textit{wudhu}.

Second, the effectiveness of the application can also be analyzed from the perspective of cognitive load. By displaying 3D animations that visually demonstrate each step, children's cognitive load can be reduced because they do not need to perform excessive abstraction processes from 2D images or verbal explanations. Children only need to observe and imitate, making the understanding process more efficient. This shows that AR applications are not only attractive but also pedagogically relevant for procedure-based learning.

Third, from a technical standpoint, the use of marker-based AR has been proven to provide adequate stability for children's learning environments. However, the use of markers also has limitations, such as dependence on lighting and camera angles. Some children have reportedly had to adjust the position of the marker card for the animation to appear correctly. This is a consideration for future development, for example by adopting markerless AR, which is more flexible and does not require additional physical media.

In addition, the 3D character models used are quite representative, but there is still room for improvement, especially in terms of facial expressions, flexibility of movement, and anatomical details, in order to make the learning experience more realistic. These visual improvements are important because young children are very sensitive to visual details and tend to be more attracted to character models that have lively expressions and are emotionally appealing.
\begin{figure}[htb!]
        \centering
        \includegraphics[width=0.5\linewidth]{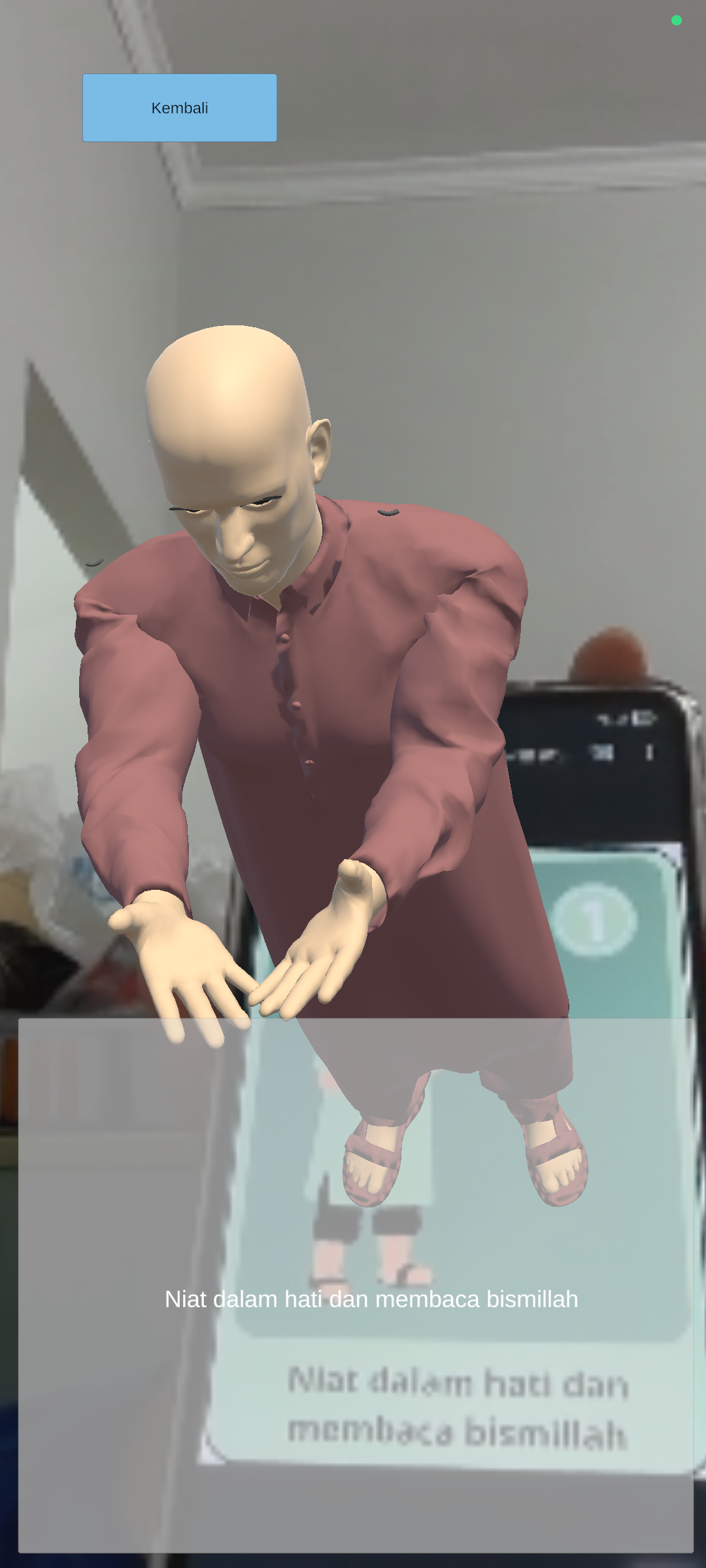}
        \caption{Augmented Reality Implementation}
        \label{fig:implementation AR}
    \end{figure}

From the perspective of teachers and assistants, this application is considered very helpful in the process of delivering material because teachers no longer have to repeat demonstrations. This shows that AR can reduce teacher's workload and provide a consistent, accurate, and replayable alternative medium. However, basic training is still needed for teachers so that they can maximize the potential of the application, especially in terms of assisting students when using the device.

Limited testing also indicates an increase in comprehension scores on the post-test compared to the pre-test, but the relatively small number of participants means that these findings cannot be generalized widely. Further research with a larger sample size and different age groups could provide a more comprehensive picture of the effectiveness of the application in the context of early childhood education.

Finally, this research opens up opportunities for developing new features such as reward systems, interactive quizzes, automatic assessment, or integration with the PAI curriculum. The addition of gamification elements, such as stars, points, or level progress, is predicted to increase children's intrinsic motivation and extend their engagement time in the learning process. Thus, this application has great potential to be developed into a more comprehensive worship learning ecosystem. As sn alternative media that is consistent, accurate, and can always be replayed. However, basic training is still needed for teachers so that they can maximize the potential of the application, especially in terms of assisting students when using the device. Limited testing also indicates an increase in comprehension scores on the post-test compared to the pre-test, but the relatively small number of participants means that these findings cannot be generalized widely. Further research with a larger sample size and different age groups could provide a more comprehensive picture of the effectiveness of the application in the context of early childhood education.

\section{Conclusion} \label{sec:conclusion}
Based on the results of implementation and testing that have been carried out, it can be concluded that the Augmented Reality (AR)-based \textit{wudhu} guidance application has been successfully developed and functions well in accordance with the research objectives. This application is capable of providing interactive learning media that combines visual, audio, and interactive elements to help early childhood understand the stages of \textit{wudhu} thoroughly.

The application of AR technology in this application has proven to be effective in presenting procedural learning materials in a more concrete and interesting way. Three-dimensional visualization and voice narration provide a more meaningful learning experience for children, in line with their cognitive development characteristics, which are still at the concrete thinking stage. From a technical perspective, the test results show that the application runs stably on mobile devices with good marker detection and an easy-to-use interface. This application is also considered to have the potential to be integrated as an auxiliary medium in teaching and learning activities in early childhood education institutions.

As a follow-up, application development can be directed towards adding other learning features such as \textit{tayamum}, prayers after \textit{wudhu}, or quiz-based game modes to make it more interesting and varied. In addition, further research involving effectiveness tests on improving children's understanding can be conducted to strengthen the empirical results of using Augmented Reality-based media in Islamic education. Overall, this study shows that the integration of Augmented Reality technology in religious education can bring about educational innovations that are relevant to the times and make a real contribution to creating a more interactive, interesting, and meaningful learning experience for early childhood.

\section*{Acknowledgment}

The author would like to thank the Faculty of Science and Technology, UIN Sunan Gunung Djati Bandung, for its support in providing facilities and guidance in conducting this research. We also extend our gratitude to the Department of Informatics for the supportive academic environment and to our colleagues who provided constructive suggestions during the development process.

%\begin{thebibliography}{00}
\bibliographystyle{./IEEEtran}
\bibliography{./IEEEabrv,./IEEEkelompok1}

%\end{thebibliography}

\end{document}